\documentclass[a4paper]{jpconf}
\usepackage{graphicx}
\usepackage{iopams}

\newcommand{\be}{\begin{eqnarray}}
\newcommand{\ee}{\end{eqnarray}}

\newtheorem{theorem}{Theorem}{}
{}
\newtheorem{lemma}{Lemma}{}
{}
\newtheorem{definition}{Definition}{}

\def\beginproof{\par\vskip 8pt\noindent{\bf Proof}\par}
\def\beginprooftheorem{\par\vskip 8pt\noindent{\bf Proof of the theorem}\par}
\def\endproof{\par\strut\hfill$\square$\par\vskip 0.5cm}

\def\Io{{\mathbb I}}
\def\Zo{{\mathbb Z}}

\begin{document}
\title{A multiplet analysis of spectra in the presence of broken symmetries}

\author{Jan Naudts$^1$ and Tobias Verhulst$^{1,2}$}

\address{
1 Departement Fysica, Universiteit Antwerpen, Universiteitsplein 1, 2610 Antwerpen, Belgium
}
\address{
2 Research Assistant of the Research Foundation - Flanders (FWO - Vlaanderen)
}

\ead{jan.naudts@ua.ac.be,tobias.verhulst@ua.ac.be}

\begin{abstract}
We introduce the notion of a generalised symmetry $M$ of a hamiltonian $H$.
It is a symmetry which has been broken in a very specific manner, involving ladder operators $R$ and $R^\dagger$.
In Theorem 1 these generalised symmetries are characterised in terms of repeated commutators of $H$ with $M$.
Breaking supersymmetry by adding a term linear in the supercharges is discussed as a motivating example.

The complex parameter $\gamma$ which appears in the definition of a generalised symmetry is necessarily real when the spectrum of $M$
is discrete. Theorem 2 shows that $\gamma$ must also be real when the spectrum of $H$ is fully discrete and
$R$ and $R^\dagger$ are bounded operators.

Any generalised symmetry induces a partitioning of the spectrum of $H$ in what we call $M$-multiplets.
The hydrogen atom in the presence of a symmetry breaking external field is discussed as an example.
The notion of stability of eigenvectors of $H$ relative to the generalised symmetry $M$ is discussed.
A characterisation of stable eigenvectors is given in Theorem 3.
\end{abstract}

\section{Introduction}

In a series of papers \cite {FSdB03,HL10,FH10,HS10,HL11,FH11} the occurrence of supersymmetry in some lattice models
has been investigated. From these studies it is clear that supersymmetry is a rather exceptional phenomenon not present in the more
common models of solid state physics. On the other hand it was pointed out in \cite {AVN08,NVA09} that many hamiltonians $H$ satisfy a
higher order commutator relation with respect to certain hermitian operators $M$. In the present paper such an operator $M$
is called a generalised symmetry of the hamiltonian $H$. This is motivated by showing that when supersymmetry is broken by adding a
perturbation term linear in the supercharges then some symmetry operators become a generalised symmetry of the new hamiltonian.

In \cite {AVN08,NVA09} the ladder operator $R$ and its conjugate $R^\dagger$ are defined as operators satisfying $[R,M]=\gamma R$.
The parameter $\gamma$ is assumed to be real. In the present work complex values of $\gamma$ are allowed. But one
of the results which we show below is that for the study of discrete spectra real-valued $\gamma$ are of prime importance.

Some of our examples involve unbounded operators. The related problems of the domain of definition of these operators are not discussed.
We use the notations
\be
[H,M]_n\equiv \underbrace{[\cdots [}_n H,M], \cdots M].
\ee

The next Section gives the definition of a generalised symmetry and proves the characterisation theorem.
Section \ref {sect:motivation} motivates our approach from the point of view of supersymmetry.
Section \ref {section:spectrum} discusses discrete spectra and their decomposition into $M$-multiplets.
The final Section contains a summary of the results.

\section{Definition and characterisation}

\subsection{Definition}
We consider triples $(H_0,M,R)$ consisting of a hamiltonian $H_0$, a symmetry $M$, and a ladder operator $R$.
They satisfy
\be
&\bullet& H_0=H_0^\dagger\mbox{ and }M=M^\dagger;\cr
&\bullet& [H_0,M]=0;\cr
&\bullet& \mbox{There exists a complex number $\gamma\not=0$ such that }[R,M]=\gamma R.\cr
& &
\label {def:triple}
\ee
Note that if $M$ has a discrete spectrum then
$\gamma$ can be taken real without restriction. Indeed, if $\psi$ is an eigenvector of $M$
with eigenvalue $m$ then either $R\psi=0$ or $R\psi$ is an eigenvector of $M$ with eigenvalue $m-\gamma$.
But the latter is real because $M$ is hermitian. One concludes that either $R=0$ or $\gamma$ is real.

An example with complex $\gamma$ can be constructed in terms of the position and momentum operators $\hat q$ and $\hat p$ of
quantum mechanics. They satisfy the commutation relations $[\hat q,\hat p]=i\hbar$ where $\hbar$ is Planck's constant.
Hence, the choices $M=\hat p\hat q+\hat q\hat p$ and $R=\hat q$ yield $[R,M]=2i\hbar R$. Further take $H_0=M$ to obtain a triple
$(H_0,M,R)$ satisfying (\ref {def:triple}).

\begin{definition}
\label {definition:gensym}
An operator $M$ is a {\sl generalised symmetry} of the hamiltonian $H$
if the latter can be written as $H=H_0+R+R^\dagger$, and the
triple $(H_0,M,R)$, satisfies the conditions (\ref {def:triple}).

\end{definition}

It is quite common to consider as a symmetry of $H$ any operator $M$ commuting with $H$.
These symmetries form a von Neumann algebra which is generated by its hermitian elements.
The justification of calling $M$ a symmetry is that when $\psi$ is an eigenvector of $H$ with eigenvalue $E$ then 
either $M\psi=0$ or $M\psi$ is again an eigenvector with the same eigenvalue $E$.
The present generalisation allows that the symmetry is broken by
adding a symmetry breaking term to the hamiltonian.

Note that the above definition is linear in $R$.
Hence varying the strength of the symmetry breaking term does not change the generalised symmetries of the hamiltonian.
Note further that $[R,M]=\gamma R$ implies that $[R^n,M]=n\gamma R^n$.
Hence the generalised symmetries of $H_0+R+R^\dagger$ are also generalised symmetries of
$H_0+\overline zR^n+z (R^n)^\dagger$, with complex $z\not=0$, and $n=1,2,\cdots$.

A point of discussion is whether one should add in the above definition the requirement that the
operators $R^\dagger R$ and $RR^\dagger$ commute with $H_0$. This is the case in many examples
and has been used in \cite {AVN08} to simplify the analysis of the spectrum of model hamiltonians.

\subsection{Commutator relations}

Generalised symmetries $M$ of a hamiltonian $H$
can be characterised by relations between higher order commutators of $H$ with $M$.

\begin{theorem}
\label {def:theo1}
A hermitian operator $M$ is a generalised symmetry of the hermitian hamiltonian $H$
if and only if one of the following cases occurs.
\begin{enumerate}
\item
 There exist a real number $\gamma_2$ with $\gamma_2\not=0$ such that
\be
\,[H,M]_2=i\gamma_2[H,M].
\label {def:tempexc}
\ee
\item 
 There exists real numbers $\gamma_1\not=0,\gamma_2$ for which
\be
[H,M]_3&=&2i\gamma_2[H,M]_2+(\gamma_1^2+\gamma_2^2)[H,M].
\label {def:3comrel}
\ee 

\end{enumerate}
\end{theorem}

Note that any genuine symmetry $M$ of $H$, this is $[H,M]=0$, satisfies (\ref {def:tempexc}) with $\gamma_2$ arbitrary
and that (\ref {def:tempexc}) implies (\ref {def:3comrel}) with $\gamma_1=0$.

\beginprooftheorem
First assume that $M$ is a generalised symmetry of the hamiltonian $H$.

Let $\gamma=\gamma_1+i\gamma_2$.
If $\gamma_1=0$ then one can derive the following equations
\be
\,H&=&H_0+R+R^\dagger
\label {def:temp0e}\\
\,[H,M]&=&i\gamma_2(R+R^\dagger)
\label {def:temp1e}\\
\,[H,M]_2&=&-\gamma_2^2 (R+R^\dagger).
\label {def:temp2e}
\ee
They imply (\ref {def:tempexc}).

Next assume that $\gamma_1\not=0$.
A straightforward calculation gives
\be
\,H&=&H_0+R+R^\dagger
\label {def:temp0}\\
\,[H,M]&=&\gamma R-\overline\gamma R^\dagger
\label {def:temp1}\\
\,[H,M]_2&=&\gamma^2 R+(\overline\gamma)^2 R^\dagger
\label {def:temp2}\\
\,[H,M]_3&=&\gamma^3 R-(\overline\gamma)^3 R^\dagger.
\label {def:temp3}
\ee
The set of equations (\ref {def:temp0},\ref {def:temp1}, \ref {def:temp2}) in
the variables $H_0$ and $R$ has the solution
\be
R&=&\frac {\overline\gamma}{2\gamma_1 |\gamma|^2}
\left([H,M]_2+\overline\gamma [H,M]\right)
\label {def:tempr}\\
H_0&=&\frac 1{|\gamma|^2}\left(-[H,M]_2+2i\gamma_2[H,M]+|\gamma|^2H\right).
\label {def:temph0}
\ee
Taking the commutator of (\ref {def:temph0}) with $M$ yields (\ref {def:3comrel}).
This ends the first part of the proof.

Next, let us prove the theorem in the other direction.
First assume (\ref {def:tempexc}).
Let
\be
H_0&=&\frac i{\gamma_2}[H,M]+H,\cr
R=R^\dagger&=&-\frac i{2\gamma_2}[H,M].
\label {def:case2expr}
\ee
They satisfy $H=H_0+R+R^\dagger$. One verifies that
\be
[H_0,M]&=&\frac i{\gamma_2} [H,M]_2+[H,M]\cr
&=&0.
\ee
Similarly is
\be
[R,M]&=&-\frac i{2\gamma_2}[H,M]_2\cr
&=&\frac 12[H,M]\cr
&=&i\gamma_2 R.
\ee
This proves the first case.

Finally assume that (\ref {def:3comrel}) holds.
Then operators $R$ and $H_0$ can be defined by
(\ref {def:tempr}) and (\ref {def:temph0}), respectively.
One verifies readily that they satisfy $H=H_0+R+R^\dagger$.
Using (\ref {def:3comrel}) one obtains
\be
[H_0,M]&=&\frac 1{|\gamma|^2}\left(-[H,M]_3+2i\gamma_2[H,M]_2+|\gamma|^2[H,M]\right)\cr
&=&0.
\ee
Similarly is
\be
[R,M]&=&\frac {\overline\gamma}{2\gamma_1 {|\gamma|^2}}
\left([H,M]_3+\overline\gamma[H,M]_2\right)\cr
&=&\gamma R.
\ee
This proves the last case.

\endproof

From the proof of the Theorem it is also clear that the operators $R,R^\dagger$ and $H_0$
are uniquely determined by the generalised symmetry $M$.

A special case of a generalised symmetry occurs when $\gamma$ is
real and different from zero.
As noted before, this is the only possibility when the generalised symmetry $M$
has a fully discrete spectrum.
Expression (\ref {def:3comrel}) becomes
\be
[H,M]_3=\gamma^2[H,M].
\label {def:3comreleal}
\ee
This special case has been investigated in \cite {AVN08,NVA09}.
In a slightly different context (\ref {def:3comreleal}) is called
the Dolan-Grady condition \cite {DG82}.

Let us consider again the quantum mechanical example.
The hamiltonian of a non-relativistic free particle with mass $m$ reads
\be
H=\frac 1{2m}\hat p^2.
\ee
Let $M=\hat p\hat q+\hat q\hat p$ as before.
Then one has $[H,M]=-4i\hbar H$. Hence, (\ref {def:tempexc}) is satisfied with $\gamma_2=-4\hbar$.
One finds $R=R^\dagger=\frac 12H$ and $H_0=0$.
This example shows that the first case of the Theorem contains generalised symmetries which are not
genuine symmetries.
Note that $M=\hat p\hat q+\hat q\hat p$ is not a generalised symmetry of the hamiltonian of the harmonic oscillator,
\be
H=\frac 1{2m}\hat p^2+\frac 12m\omega_0^2\hat q^2
\ee
because $[H,M]_3=-\gamma^2[H,M]$ holds with $\gamma=4\hbar$.
This relation has the wrong sign to be written in the form (\ref {def:3comreleal}).
This would mean that $iM$ is a generalised symmetry.
However, in the present work we require generalised symmetries to be hermitian.

Examples of the second case of the Theorem are discussed in the next Section.

\subsection{Existence}
\label {def:subsect:existence}

An obvious question is whether there exist always generalised symmetries of
a given hermitian hamiltonian $H$, other than the genuine symmetries. The answer is positive
since all orthogonal projection operators are generalised symmetries of any hermitian hamiltonian.
Indeed, if $M^2=M$ then one has
\be
[H,M]_3=HM^3-3MHM^2+3M^2HM-M^3H=[H,M].
\ee
Hence, by Theorem \ref {def:theo1}, $M$ is a generalised symmetry of $H$.
Similarly, if $M^2=\Io$ then
\be
[H,M]_3=HM^3-3MHM^2+3M^2HM-M^3H=4[H,M].
\ee
Hence orthogonal $M$ are generalised symmetries as well.
If one would add to the Definition \ref {definition:gensym} the requirement that
$M$ commutes with $R^\dagger R$ and $RR^\dagger$ then these examples do not work in general.
In any case, due to the nonlinear nature of the commutation relations it is unclear what is the
structure of the set of generalised symmetries.

\section{Motivation}
\label {sect:motivation}

The main motivation for introducing the notion of generalised symmetries is
that the operators $R$ and $R^\dagger$ can act as ladder operators --- see the next Section.
But its origin lies in the study of model hamiltonians and in an effort to generalise
the notion of supersymmetry. The idea is that even when a symmetry is broken
some traces of this symmetry may remain. The generalised symmetry $M$ of $H$  is a genuine
symmetry of the hamiltonian $H_0$. The symmetry breaking term $R+R^\dagger$ has
special properties which can be used to relate the spectrum of $H$ to that of $H_0$.

\subsection{Lie superalgebras}

The simplest non-trivial Lie superalgebra contains two odd generators $Q$ and $Q^\dagger$
and one even generator $\{Q,Q^\dagger\}$. The latter commutes with all elements of the
algebra and is for that reason called supersymmetric. Consider now the hamiltonian
\be
H=\{Q,Q^\dagger\}+\overline z Q+zQ^\dagger,
\ee
with complex $z$. For $z\not=0$ the supersymmetry is broken. But the hamiltonian still belongs
to the Lie superalgebra.

Let us now add to the  Lie superalgebra a second even generator, denoted $M$.
The following result is an immediate consequence of the assumption that the new algebra has exactly 4 generators
and that we assume a representation as operators in a Hilbert space, with $Q^\dagger$ the adjoint of
$Q$ and with $M=M^\dagger$.

\begin{lemma} There exists a complex number $\gamma$ such that
\be
[Q,M]=\gamma Q.
\label {super:gamma}
\ee
\end{lemma}

\beginproof

Because $M$ is even and $Q$ is odd also $[Q,M]$ is odd. Hence it can be written as
\be
[Q,M]=\gamma Q+\xi Q^\dagger
\ee
with $\gamma$ and $\xi$ complex numbers.
Using $Q^2=0$, multiplying once from the left and once from the right, one obtains
\be
-QMQ&=&\xi QQ^\dagger\cr
QMQ&=&\xi Q^\dagger Q.
\ee
Adding the two together gives $0=\xi \{Q,Q^\dagger\}$. But $\{Q,Q^\dagger\}$ is an even generator of the
superalgebra. Hence it cannot vanish. Therefore one concludes that $\xi=0$.

\endproof

Now write $H=H_0+R+R^\dagger$ with $H_0=\{Q,Q^\dagger\}$ and $R=\overline z Q$.
The previous Lemma yields $[R,M]=\gamma R$. Further is
\be
[H_0,M]=[QQ^\dagger+Q^\dagger Q,M]=0
\ee
Hence, $M$ is a generalised symmetry of the hamiltonian $H$.
Note that $M$ commutes with $Q^\dagger Q$ and $QQ^\dagger$ as well.

In 1981 Witten \cite {WE81} introduced his non-relativistic model of supersymmetry.
The one-particle hamiltonian reads
\be
H_0=-\frac {\hbar^2}{2m}\frac {{\rm d}^2\,}{{\rm d}q^2}+\frac 12mw^2(q)
+\frac 12\hbar \sigma_z\frac {{\rm d}w}{{\rm d}q}.
\label {SUSYham}
\ee
It can be written as
\be
H_0=\{Q,Q^\dagger\}
\label {superchargerel}
\ee
with
\be
Q=\frac 1{\sqrt{2m}}\left(\hat p-im w(q)\right)\sigma_+,
\ee
where
\be
\hat p=\frac {\hbar}i\frac {{\rm d}\,}{{\rm d}q}
\quad\mbox{ and }\quad
\sigma_\pm=\frac 12(\sigma_x\pm i\sigma_y).
\ee
($\sigma_\alpha,\alpha=x,y,z$ are the Pauli matrices).

The Pauli matrix $M=\sigma_z$ can be added as a fourth generator to the super algebra
generated by $Q,Q^\dagger,H_0$. It satisfies $[Q,M]=2Q$. This illustrates the previous Lemma.
That $M$ is a generalised symmetry of $H=H_0+\overline zQ+zQ^\dagger$
is a rather trivial result because $M^2=\Io$ --- see the remark in subsection \ref {def:subsect:existence}.
A short calculation shows that $Q^\dagger Q$ and $QQ^\dagger$ commute with $M$. 

\subsection{Lattice models}

Let $B_j,B_j^\dagger$ denote the annihilation and creation operators of a fermionic particle at the lattice site $j$
belonging to a finite subset $\Lambda$ of $\Zo^n$. They satisfy $\{B_j,B_k\}=\delta_{j,k}$
and are the odd generators of a Lie superalgebra with $2n$ odd generators and one even generator, which is the identity
operator. The latter is supersymmetric. The physically interesting hamiltonian is
\be
H_0=-\epsilon\sum_{i,j\in\Lambda}^{|i-j|=1}B^\dagger_iB_j.
\label {mot:hop}
\ee
It describes the hopping of fermions between neighbouring places. $\epsilon$ is a numerical constant.
One has
\be
[H_0,B_i]=\epsilon\sum_j^{|i-j|=1}B_j.
\ee
Hence, $H_0$ may be added to the super algebra as n even generator.
Of further interest is the operator
\be
M=\sum_{i\in\Lambda}B^\dagger_iB_i.
\ee
It counts the number of fermions and it satisfies $[B_j,M]=B_j$ and $[H_0,M]=0$.
Hence, it can also be added as an even generator. Moreover, the triples $(H_0,M,B_i)$
satisfy the conditions (\ref {def:triple}) with $\gamma=1$.
Hence $M$ is a generalised symmetry of each of the hamiltonians 
$H_0+\overline z_iB_i+z_iB_i^\dagger$, where $z_i$ is a complex constant.
It is also a generalised symmetry of the hamiltonian
\be
H=H_0+\sum_i\left(\overline z_iB_i+z_iB_i^\dagger\right).
\ee
However, this hamiltonian is physically not so interesting because fermions are
usually conserved in number, while now this symmetry is broken. But by taking $z_i=0$ except at
the borders one can model the effect that fermions leave or enter the system.

\subsection{Hard core repulsion}

Following \cite {FSdB03, HL10,HS10,HL11}
the model hamiltonian (\ref {mot:hop}) can be made more interesting by including a nearest neighbour exclusion
mechanism. Let
\be
Q=\sum_i P_iB_i^\dagger
\quad\mbox{ with }\quad
P_i=\prod_j^{|i-j|=1}B_jB_j^\dagger.
\ee
Note that $Q^2=0$.
A supersymmetric hamiltonian is given by
\be
H_0=\{Q,Q^\dagger\}
&=&\sum_{i,j\in\Lambda}\left(P_iB_i^\dagger B_jP_j+B_jP_jP_iB_i^\dagger\right)\cr
&=&\sum_{i\in\Lambda}P_i
+\sum_{i,j\in\Lambda}^{|i-j|=1}P_iB_i^\dagger B_jP_j.
\ee
The last term describes a hopping of spinless electrons between neighbouring sites under the
condition that never two electrons occupy neighbouring positions. The first term counts the numbers of sites
which have all neighbouring sites unoccupied. The perturbation term
\be
\overline z Q+zQ^\dagger=\sum_iP_i(\overline z B_i^\dagger+zB_i) 
\label {schout:source}
\ee
is an external field injecting or removing electrons at sites where they do not have neighbours.

The counting operator $M=\sum_i B_i^\dagger B_i$ is a generalised symmetry of this model.
Indeed, one has $[Q,M]=-Q$. By adding the source term (\ref {schout:source}) the conservation
of the number of electrons is broken. But it remains a generalised symmetry.

\subsection{Jaynes-Cummings model}

Another way of making (\ref {mot:hop}) more interesting is by replacing the symmetry-breaking terms
$\overline z Q+zQ^\dagger$ by an interaction with a bosonic particle.
The prototype model is that of Jaynes and Cummings.
The hamiltonian is, see for instance \cite {RJC99},
\be
H=\frac 12\hbar\omega_0\sigma_z+\frac 12\hbar\omega (c^\dagger c+cc^\dagger)
+\hbar\kappa(c^\dagger\sigma_-+c\sigma_+),
\ee
where $c,c^\dagger$ are the annihilation and creation operators of a harmonic oscillator. They satisfy $[c,c^\dagger]=\Io$.
The $\sigma_\alpha$ are the Pauli matrices and $\sigma_\pm=\frac12(\sigma_x\pm i\sigma_y)$.
They satisfy $\{\sigma_+,\sigma_-\}=\Io$.
The model is said to be at resonance when $\omega=\omega_0$. 
When at resonance and without interaction term (this is, $\kappa=0$)
the model is supersymmetric with $Q=\sqrt{\hbar\omega} c^\dagger\sigma_-$
(neglecting a shift in the energy scale).
The symmetry breaking term $\overline z Q+zQ^\dagger$ reproduces the interaction term.
Indeed, one obtains
\be
H_*=\{Q,Q^\dagger\}+ z Q+zQ^\dagger=\hbar\omega\sigma_+\sigma_-+\hbar\omega c^\dagger c
+\hbar\kappa(c^\dagger\sigma_-+c\sigma_+),
\ee
with $z=\hbar\kappa/\sqrt{\hbar\omega}$.
Note that $[Q,\sigma_z]=-Q$ and $[\{Q,Q^\dagger\},\sigma_z]=0$. Hence $\sigma_z$ is a generalised
symmetry of $H_*$.
Off resonance the model is not anymore supersymmetric.
But even then $\sigma_z$ remains a generalised symmetry of $H$.
Indeed, one has
\be
[H,\sigma_z]&=&\hbar\kappa(c^\dagger\sigma_--c\sigma_+)\cr
[H,\sigma_z]_2&=&\hbar\kappa(c^\dagger\sigma_-+c\sigma_+)\cr
[H,\sigma_z]_3&=&\hbar\kappa(c^\dagger\sigma_--c\sigma_+)=[H,\sigma_z].
\ee
By Theorem \ref {def:theo1} this implies that $\sigma_z$ is a generalised symmetry of $H$.
One finds $\gamma=1$ and $R=\hbar\kappa c^\dagger\sigma_-$. Note that $R^\dagger R$ and $RR^\dagger$
both commute with $\sigma_z$.

\section{The spectrum of $H$}
\label {section:spectrum}

If $M=M^*$ is a genuine symmetry of the hamiltonian $H$ then the unitary transformations
\be
A\rightarrow e^{-i\lambda M}A e^{i\lambda M}
\ee
leave the hamiltonian invariant. If $M$ is a generalised symmetry then one has
\be
e^{-zM}He^{zM}=H_0+e^{z\gamma}R+e^{-z\overline\gamma}R^{\dagger}.
\label {main:simul}
\ee
This similarity transformation does not change the spectrum of the hamiltonian.
But it modulates the symmetry breaking term with a complex factor.
The eigenvectors transform as $\psi\rightarrow e^{-zM}\psi$.
In what follows we explore whether this relation creates a link between eigenvalues of $H_0$ and of $H$.

\subsection{Eigenvectors}

The following result suggests that for the study of discrete spectra of hamiltonians only
generalised symmetries with a real $\gamma$ are of interest.
This is the case which has been studied in \cite {AVN08,NVA09}.

\begin{theorem}
\label {spectrum:theo2}
 Let $M=M^\dagger$ be a generalised symmetry of the hamiltonian $H=H^\dagger$, with associated triple
$(H_0,M,\gamma)$. Assume that $H$ has a fully discrete spectrum and that $M$ is not a symmetry of $H$.
Assume in addition that $R$ is a bounded operator.
Then $\gamma$ is necessarily real.
\end{theorem}

In particular, if $H$ has a fully discrete spectrum and $R$ is a bounded operator
then case 1 of Theorem \ref {def:theo1} only contains genuine symmetries.
It is not clear whether the condition that $R$ is a bounded operator can be removed.

\beginprooftheorem
Let us assume that $\Im\gamma\not=0$.
Let $\psi$ be a normalised eigenvector of $H$, with eigenvalue $E$.
\be
\psi_\alpha=e^{i\alpha M}\psi
\quad\mbox{ with }\alpha \mbox{ real}.
\ee
Then $||\psi_\alpha||=1$.
One calculates using (\ref {main:simul})
\be
||(H_0-E)\psi||=||(e^{i\alpha\gamma}R+e^{-i\alpha\overline\gamma}R^{\dagger})\psi_\alpha||
\le 2e^{-\alpha\Im\gamma}||R||.
\ee
Depending on the sign of $\Im\gamma$ one lets tend $\alpha$ to $+\infty$ or $-\infty$
and one concludes that $\psi$ is an eigenvector of $H_0$ with eigenvalue $E$.
Because the spectrum of $H$ is fully discrete this implies that $H$ and $H_0$ coincide
and hence that $H$ commutes with $M$. But by assumption $M$ is not a genuine symmetry of $H$.
Hence, the assumption that $\Im\gamma\not=0$ is false.
One concludes that $\gamma$ is real.

\endproof

\subsection{The multiplet structure}

When breaking a symmetry of a hamiltonian degenerate energy levels will split up.
After breaking the symmetry the
corresponding eigenvectors still transform into each other under the symmetry operations of the
original hamiltonian. They form a so-called multiplet. This traditional way of defining multiplets of eigenvalues
is now adapted. Starting point is the observation that the definition of a multiplet depends
on the broken symmetry under consideration.

\begin{definition}
\label {multiplet:def}
Let $M$ be a generalised symmetry of the hamiltonian $H$. Two eigenvectors $\phi$ and $\psi$ of $H$
are said to belong
to the same $M$-{\sl multiplet} if there exists an invertible function $f$ such that $\phi=f(M)\psi$.
\end{definition}

The above definition is an extension of the existing notion of multiplets.
In \cite {NVA09} the example of the harmonic oscillator is considered with the shifted harmonic oscillator
as the generalised symmetry $M$. Then all eigenvalues belong to the same $M$-multiplet. For instance, the first excited state
$|1\rangle$ can be written as $|1\rangle=f(M)|0\rangle$, where $|0\rangle$ is the ground state and
$f(x)$ is a linear function.
 
Note that the similarity transformation (\ref {main:simul}) maps $M$-multiplets onto themselves.

\subsection{The hydrogen atom in an external potential}
\label {subsection:hydrogen}

An obvious example is the hamiltonian  $H_0$ of the hydrogen atom. It is given by
\be
H_0=\frac 1{2m}\sum_\alpha \hat p_\alpha^2-\frac {e^2}{\sqrt{\sum_\alpha \hat q_\alpha^2}},
\ee
where $e,m$ are constants and $\hat q_\alpha,\hat p_\alpha,\alpha=1,2,3$ are the position and momentum operators of the electron.
The hamiltonian is invariant under spatial rotations. Therefore the components $L_\alpha$
of the angular momentum commute with $H_0$.

Now add to this hamiltonian a symmetry breaking potential.
The simplest one is a uniform field in the $x$-direction
\be
H=H_0-2g\hat q_x
\ee
with real coupling constant $g$.
This hamiltonian can be written in the form $H=H_0+R+R^\dagger$,
with $R=-g(\hat q_x+i\hat q_y)$.
From the commutation relations
\be
[L_z,\hat q_x]=i\hbar \hat q_y,\quad
[L_z,\hat q_y]=-i\hbar \hat q_x,\quad
[L_z,\hat q_z]=0
\ee
follows immediately that $M=L_z$ is a generalized symmetry of $H$, with $\gamma=-\hbar$.

The method of classifying eigenvalues into multiplets, as worked out in \cite {AVN08},
works well if a basis of common eigenvectors of $H_0$ and $M$ is explicitly known.
This is only partially the case for the hamiltonian of the hydrogen atom
with $M=L_z$ since the spectrum of these operators is partly discrete, partly continuous.
In addition, the perturbed hamiltonian $H$ introduced above does not
have any discrete eigenvalue. Therefore it is meaningless to discuss its multiplet structure.
Let us therefore look for an alternative symmetry breaking potential.

A basis of eigenvectors of the hydrogen atom can be labeled with quantum numbers $n,l,m$
with $n=1,2,\cdots$, $l=0,1,\cdots,n-1$, $m=-l,-l+1,\cdots,l$. They satisfy
\be
H_0|n,l,m\rangle&=&E_n|n,l,m\rangle,\cr
\sum_{\alpha}L_\alpha^2|n,l,m\rangle&=&\hbar l(l+1)|n,l,m\rangle,\cr
L_z|n,l,m\rangle&=&\hbar m|n,l,m\rangle,
\ee
with $E_n$ proportional to $-1/n^2$.
The conventional ladder operators are $L_\pm=L_x\pm iL_y$. They satisfy
\be
L_\pm|n,l,m\rangle\sim |n,l,m\pm 1\rangle.
\ee
Consider therefore the hamiltonian
\be
H=H_0-2g L_x=H_0+R+R^\dagger
\quad\mbox{ with }R=-gL_-.
\ee
Then $M=L_z$ is a generalised symmetry of $H$ with $\gamma=-\hbar$.
Note that in this case $R$ commutes with $H_0$. Hence the Hilbert space can be restricted to the subspace
generated by the eigenvectors of $H_0$. The eigenvectors of $H$ form multiplets indexed
by the quantum numbers $n,l$. Within such a multiplet the relevant operators can be represented as
square matrices of size $2l+1$. In this context any diagonal matrix can be written as
a function of $M=L_z$. Therefore all wavefunctions of the form
\be
\psi=\sum_{m=-l}^lc_m|n,l,m\rangle
\ee
for which none of the $c_m$ vanish belong to the same $M$-multiplet.
But the multiplet with quantum numbers $n,l$ with $l\ge 1$ always contains at least one eigenvector $\psi$ of $H$ which satisfies
$c_{-m}=-c_m$ and hence $c_0=0$ --- see below. This makes clear that the multiplet with quantum numbers $n,l$
decomposes into more than one $M$-multiplet.

The requirement that $H\psi=E\psi$ leads to the set of equations
\be
Ec_m&=&E_nc_m+c_{m+1}\langle m|R|m+1\rangle+c_{m-1}\langle m|R^\dagger|m-1\rangle.
\label {hydro:seq}
\ee
The matrix elements of $R$ and $R^\dagger$ are given by (see for instance Section VI.40 of \cite {DAS65})
\be
\langle m+1|R^\dagger|m\rangle=\langle m|R|m+1\rangle
=-g\frac {\hbar}{\sqrt 2}\,\sqrt{(l-m)(l+m+1)}
\ee
The {\sl ansatz} that $c_0=0$ implies $c_{-1}=-c_1$.
It is then easy to see that the equations can be solved iteratively, maintaining $c_{-m}=-c_m$.
The top equation
\be
Ec_l&=&E_nc_l+c_{l-1}\langle l|R^\dagger|l-1\rangle
\ee
then fixes the eigenvalue $E$.
If $l=1$ this forces $E=E_n$ and $(R+R^\dagger)\psi=0$.
The $l=1$ triplets therefore decompose into an $M$-singlet and an $M$-doublet.

The $l=2$ multiplets can be shown to decompose into two $M$-doublets
and one $M$-singlet with eigenvector of the form $(x,0,z,0,x)^{\rm T}$. In general, the decomposition into $M$-multiplets
is governed by the occurrence of vanishing coefficients $c_m$.

\subsection{Stable eigenvectors}

The following definition is taken from \cite {NVA09}.

\begin{definition}
 The eigenvector $\psi$ of the hamiltonian $H$ is {\sl stable} relative to the
generalised symmetry $M$ if the vectors $\psi, R\psi,R^\dagger\psi$ are linearly dependent.
\end{definition}

Eigenvectors $\psi$ for which $R\psi=0$ or $R^\dagger\psi=0$ are clearly stable.
The $l=1$ $M$-multiplets in the example of the hydrogen atom all contain a pair of stable eigenvectors ---
the proof follows later on.
This is not by accident, as shows the next theorem, which improves on a result from \cite {NVA09}.

\begin{theorem}
Let be given an eigenvector $\psi$ of a hamiltonian $H$.
Assume that $M$ is a generalised symmetry of $H$, with real $\gamma$.
Then $\psi$ is stable relative to $M$ if and only if
one of the following possibilities holds.\\
1) There exist numbers $x$ and $y$, not both zero, such that $xR\psi+yR^\dagger \psi=0$.\\
2) $\psi$ is an eigenvector of $R$.\\
3) $\psi$ is an eigenvector of $R^\dagger$.\\
4) $\psi$ is an eigenvector of $R-R^\dagger$.\\
5) There exists a complex $z$ such that $\chi=e^{-zM}\psi$ is again an eigenvector of $H$
with eigenvalue different from that of $\psi$.
\end{theorem}

\beginprooftheorem
Assume first that $\psi$ is stable relative to $M$.
Then there exist $x,y,u$ not all zero such that
\be
xR\psi+yR^\dagger\psi=u\psi.
\ee
The choice $u=0$ corresponds with the first possibility.
Let us therefore assume that $u\not=0$. Then we can assume that $u=1$ without restriction.
The possibilities $x=0$ or $y=0$ then correspond with cases (2) or (3).
Remains the situation when $xy\not=0$.
The case that $x+y=0$ leads to possibility (4) of the Theorem.
Assuming $x+y\not=0$ leads to the possibility (5).
Indeed, there exists $z$ satisfying $e^{z\gamma}=-y/x\not=1$ such that
\be
x(e^{z\gamma}-1)=y(e^{-z\gamma}-1).
\label {stable:zeq}
\ee
This leads to, assuming $H\psi=E'\psi$,
\be
H\chi&=&He^{-zM}\psi\cr
&=&(H_0+R+R^\dagger)e^{-zM}\psi\cr
&=&e^{-zM}(E'+(e^{-z\gamma}-1)R+(e^{z\gamma}-1)R^\dagger)\psi\cr
&=&e^{-zM}\left(E'+\frac 1x(e^{-z\gamma}-1)[xR+yR^\dagger]\right)\psi\cr
&=&e^{-zM}\left(E'+\frac 1x(e^{-z\gamma}-1)\right)\psi\cr
&=&E''\chi
\quad\mbox{ with }
E''=E'+\epsilon\cr
& &\qquad
\quad\mbox{ and }\epsilon=\frac 1x(e^{-z\gamma}-1)=\frac 1y(e^{z\gamma}-1).
\label {stable:temp}
\ee
Note that $\epsilon\not=0$ because of $e^{z\gamma}\not=1$.
This covers case (5) and ends the proof in one direction.

The other direction of the proof is straightforward except in the last case.
Assume that both $\psi$ and $\chi=e^{-zM}\psi$ are eigenvectors of $H$ with corresponding
eigenvalues $E',E''$.
Then one has
\be
E'\psi&=&H\psi\cr
&=&(H_0+R+R^\dagger)e^{zM}\chi\cr
&=&e^{zM}(H_0+e^{z\gamma}R+e^{-z\gamma}R^\dagger)\chi\cr
&=&e^{zM}(H+(e^{z\gamma}-1)R+(e^{-z\gamma}-1)R^\dagger)\chi\cr
&=&e^{zM}(E''+(e^{z\gamma}-1)R+(e^{-z\gamma}-1)R^\dagger)\chi\cr
&=&[E''+(1-e^{-z\gamma})R+(1-e^{z\gamma})R^\dagger]\psi.
\ee
Because $E'\not=E''$ this shows that $\psi,R\psi,R^\dagger \psi$
are linearly dependent. Hence $\psi$ is stable relative to $M$.

\endproof

The situation of a ladder operator $R$ annihilating the eigenstate $\psi$ is a subcase of
case (1) of the theorem ($x=1$ and $y=0$). It is well-known because the vacuum vector of the harmonic oscillator
is annihilated by the standard boson annihilation operator $b$. But in \cite {NVA09} the shifted harmonic
oscillator is shown to be a generalised symmetry of the harmonic oscillator. Then the operator $R$
is a linear combination of $b$ and the constant operator. The ground state is then an eigenvector of $R$.
Thus this is an example of case (2) of the theorem.

A special subcase of case (1) of the theorem is when $(R+R^\dagger)\psi=0$.
Then one has $H\psi=H_0\psi=E\psi$. Hence, these $\psi$ are simultaneous eigenvectors of $H$ and $H_0$.
Examples of such stable eigenvectors are the $l=0$ singlets of the hydrogen atom in an external field,
discussed in Subsection \ref {subsection:hydrogen},
because they are annihilated by both $R$ and $R^\dagger$.

In case (4) is $R\psi=R^\dagger\psi$.
From (\ref {def:tempr}) then follows that $[H,M]\psi=0$ (note that $\gamma$ is taken to be real).
Using $H\psi=E\psi$ then follows that $M\psi$ is again an eigenvector of $H$ also with eigenvalue $E$.
Note that $M\psi$ cannot vanish.
This is a rather pathological situation which we did not encounter yet in any example.

The last case of the theorem is the generic case. It predicts the existence of pairs of stable eigenvectors
belonging to the same $M$-multiplet. Examples have been discussed in \cite {NVA09}.
Such a pair of stable eigenvectors ia also found in the $l=1$ multiplets of the hydrogen atom in an external field
(see Subsection \ref {subsection:hydrogen}).

\section{Summary}

We study a class of symmetries $M=M^\dagger$ which are broken by adding to the
hamiltonian $H_0$ a perturbation of the form $R+R^\dagger$, where $R$ is a ladder operator satisfying
$[R,M]=\gamma R$ with complex $\gamma$.
Because the symmetry is only broken in this special way we call $M$ a {\sl generalised symmetry} of $H=H_0+R+R^\dagger$.

These generalised symmetries are characterised by a commutator relation --- see Theorem \ref {def:theo1}.
While a genuine symmetry $M$ commutes with the hamiltonian $H$ the generalised symmetry satisfies
relation (\ref {def:3comrel}) which involves also the second and third commutators of $H$ with $M$.
In the case that $\gamma$ is real this commutator relation has been studied in \cite {AVN08,NVA09}.
In a slightly different context it has
been called the Dolan-Grady relation \cite {DG82}.

Generalised symmetries, in the weak form of Definition \ref {definition:gensym},
do always exists and can be helpful in studying the discrete spectrum of quantum hamiltonians.
Several examples have been mentioned throughout the paper. Many of these examples involve the case
of $\gamma$ being real, which was already studied in \cite {AVN08,NVA09}.
This not by accident. If the generalised symmetry $M$ has a discrete spectrum then $\gamma$ must be real.
In addition, Theorem \ref {spectrum:theo2} shows that hamiltonians with a fully discrete spectrum
and satisfying some technical condition,
only have generalised symmetries with real $\gamma$. As a consequence the present generalisation from
real to complex $\gamma$ does not immediately lead to new applications.

We introduce a formal definition of $M$-multiplets based on the work of \cite {AVN08,NVA09}.
We show for the example of the hydrogen atom how this definition relates with the usual notion
of multiplets of degenerate eigenvectors.

We finally discussed the notion of eigenvectors stable relative to a generalised symmetry.
This notion was introduced in \cite {NVA09}.
Such stable eigenvectors, if known, can help to find new eigenvectors in a way reminiscent
of the Darboux method to find new solutions of nonlinear differential equations out of
known solutions.

The theory of generalised symmetries as presented here is probably not yet in its final form.
It has many applications one of which is a partition of energy spectra into multiplets
in a way depending on the choice of a generalised symmetry.
From these applications one can learn how to extend our knowledge about generalised symmetries.


\section*{References}
\begin{thebibliography}{99}

\bibitem {FSdB03}
 Fendley P, Schoutens K and de Boer J 2003
Lattice models with N = 2 supersymmetry
{\sl Phys. Rev. Lett.} {\bf 90} 120402

\bibitem {HL10}
Huijse L 2010
{\sl A supersymmetric toy model for lattice fermions}
(PhD Thesis, Universiteit van Amsterdam)

\bibitem {FH10}
Fendley P and Hagendorf Ch 2010
Exact and simple results for the XYZ and strongly interacting fermion chains
{\sl J. Phys.} A: Math. Theor. {\bf 43} 402004

\bibitem {HS10}
Huijse L and Schoutens K 2010
Supersymmetry, lattice fermions, independence complexes and cohomology theory
{\sl Adv. Theor. Math. Phys.} {\bf 14}(2) 643--694 (2010).

\bibitem {HL11}
Huijse L 2011
Detailed analysis of the continuum limit of a supersymmetric lattice model in 1D
{\sl J. Stat. Mech.}  P04004 

\bibitem {FH11}
Fendley P and Hagendorf Ch 2011
 Ground-state properties of a supersymmetric fermion chain
 {\sl J. Stat. Mech.}  P02014 


\bibitem {AVN08}
Verhulst T, Anthonis B and Naudts 2009
Analysis of the N=4 Hubbard ring using counting operators
{\sl Phys. Lett.} A {\bf 373} 2109--2113 

\bibitem {NVA09}
Naudts J, Verhulst T and Anthonis B 2009
Counting operator analysis of the discrete spectrum of some model hamiltonians
{\sl Phys. Lett.} A {\bf 373} 3419--3422 

\bibitem {DG82}
Dolan L and Grady M 1982
Conserved charges from self-duality
 {\sl Phys. Rev.} D {\bf 25} 1587--1604

\bibitem {WE81}
Witten E 1981
Dynamical breaking of supersymmetry
 {\sl Nucl. Phys.}  B {\bf 188} 513--554

\bibitem {RJC99}
Rajagopal A K, Jensen K L and Cummings F W 1999
 Quantum entangled supercorrelated states in the Jaynes-Cummings model
 {\sl Phys. Lett.} A {\bf 259} 285--290

\bibitem {DAS65}
 Davydov A S 1965
{\sl Quantum Mechanics}
(Pergamon Press) 2nd ed

\end{thebibliography}
\end{document}